\documentclass[a4paper,11pt]{article}
\usepackage{latexsym}
\usepackage{amsmath}
\usepackage{amssymb}
\usepackage{graphicx}
\usepackage{wrapfig}
\begin{document}

\baselineskip=7mm


\small

\noindent
SAGA-HE-221-05

~

\centerline{\bf Auxiliary Field Meson Model at Finite Temperature and Density
}

\small
\baselineskip=2mm

~

\centerline{H. Kouno$^1$, T. Sakaguchi$^2$, K. Kashiwa$^2$, M. Hamada$^2$, 
}
\centerline{H. Tokudome$^1$, M. Matsuzaki$^{2,3}$ and M. Yahiro$^2$}
\centerline{1 \it Department of Physics, Saga University, Saga 840-8502, Japan}
\centerline{2 \it Department of Physics, Kyushu University, Fukuoka 812-8581, Japan}
\centerline{3 \it Department of Physics, Fukuoka University of Education, }
\centerline{\it Munakata, Fukuoka 811-4192, Japan}

\baselineskip=2mm

~

\small
\centerline{\bf Abstract}
Starting from many quark interactions, we construct a nonlinear $\sigma$-$\omega$ model at finite temperature and density. 
The mesons are introduced as auxiliary fields. 
Effective quark-meson couplings are strongly related to effective meson masses, since they are derived simultaneously from the original many quark interactions. 
In this model, even if the effective $\omega$-meson mass decreases due to the partial chiral restoration, the equation of state (EOS) of nuclear matter can become soft.  

~

\noindent
{\bf 1 Introduction}

The $\omega$-meson is important for the nuclear structure. 
It is reported that the reduction of the effective $\omega$-meson mass makes the nuclear matter EOS stiffer. [1] 
However, if we require that the $\omega$-meson mean field be proportional to the baryon density, the effective $\omega$-nucleon coupling also becomes smaller as the effective $\omega$-meson mass becomes smaller and the EOS of nuclear matter becomes softer. [2] 

In this paper, we show that, at finite temperature and density, effective meson-quark couplings are strongly related to effective meson masses, if the meson fields are introduced as auxiliary fields which consist of quarks and anti-quarks. 
Consequently, if the effective $\omega$-meson mass decreases, the effective $\omega$-quark (or nucleon) coupling decreases and the EOS of nuclear matter becomes softer. 
Therefore, even if the effective $\omega$-meson mass decreases due to the partial chiral restoration, the EOS of nuclear matter can become soft in this model.   

~

\noindent
{\bf 2 Auxiliary field method for nonlinear $\sigma$-$\omega$ model}

In this section, using the auxiliary field method, [3,4] 
we construct a nonlinear $\sigma$-$\omega$ model. 
(For details, see the reference [5].) 
We start from the many quark interactions [4,5] 
\begin{eqnarray}
\int dt~V&=&\sum_{m+n^>_=2}{1\over{m!n!}}\int d^4x_1\cdots d^4x_md^4y_1\cdots d^4y_m
d^4u_1\cdots d^4u_nd^4v_1\cdots d^4v_n
\nonumber\\
&\times& V^{(m,n)}_{\mu_1,\cdots ,\mu_n}(x_1,\cdots,x_m,y_1,\cdots,y_m,
u_1\cdots u_n,v_1\cdots v_n)
\nonumber\\
&\times&:\bar{\psi}(x_1)\psi (y_1)\cdots\bar{\psi}(x_m)\psi (y_m)\bar{\psi}(u_1)\gamma^{\mu_1}\psi (v_1)\cdots\bar{\psi}\gamma^{\mu_n}\psi (v_n):, 
\label{eq:E1}
\end{eqnarray}
where $\psi$ is the quark field. 
The quantum transition amplitude is given by 
\begin{eqnarray}
Z_{\rm fi}&=&\int D\psi D\bar{\psi}\exp{\left(i\int d^4x L\right)}, 
\label{eq:E2}
\end{eqnarray}
where $L$ is the Lagrangian density of the system. 
Inserting the identity 
\begin{eqnarray}
1
&=&\int \prod_{x,y}D\Sigma_{\rm s}(x,y)D\Sigma_\mu(x,y)D\sigma (x,y)D\omega^\mu(x,y)
\nonumber\\
&&\exp{\left(i\int dxdy\Sigma_{\rm s}
\left\{\sigma (x,y)-\bar{\psi}(x)\psi (y)\right\}\right)}
\nonumber\\
&&\times\exp{\left(i\int dxdy\Sigma_\nu
\left\{\bar{\psi}(x)\gamma^\nu\psi (y)-\omega^\nu (x,y)\right\}\right)}, 
\label{eq:E4}
\end{eqnarray}
we introduce the auxiliary meson fields $\sigma (=\bar{\psi}\psi )$ and $\omega _\mu (=\bar{\psi}\gamma_\mu\psi )$ as well as the quark self-energies $\Sigma_{\rm s}$ and $\Sigma_\mu$. 
In this model, the expectation values of $\sigma$ and $\omega_0$ fields are proportional to the quark scalar density and the baryon density, respectively. 

Integrating the quark field,  
we obtain by means of the mean field approximation 
\begin{eqnarray}
Z_{\rm fi}=\int D\sigma D\omega_\mu \exp{\left(i\Gamma [\sigma,\omega_\mu ]\right)}, 
\label{eq:E16}
\end{eqnarray}
where $\Gamma$ is the effective action and is given by 
\begin{eqnarray}
\Gamma [\sigma,\omega_\mu ]
&=&W_0[\Sigma_{\rm s}[\sigma,\omega_\mu ],\Sigma_\mu [\sigma,\omega_\mu ]]
-\sum_{m+n^>_=2}\int V^{(m,n)}_{\mu_1\cdots\mu_n}\sigma^m\omega^{\mu_1}\cdots\omega^{\mu_n} 
\nonumber\\
&+&{\rm Tr}\left(\sigma\Sigma_{\rm s}[\sigma,\omega_\mu ]-\omega_\mu\Sigma^\mu [\sigma,\omega_\mu ]\right) .
\label{eq:E17}
\end{eqnarray}
The $W_0$ represents the quark energy and the remaining parts represent the meson potential. 
The quark self-energies $\Sigma_{\rm s}$ and $\Sigma_\mu$ are determined by 
the following conditions. 
\begin{eqnarray}
{\delta \Gamma\over{\delta \sigma}}
&=&-{\partial\over{\partial \sigma}}\left(\sum_{m+n^>_=2}\int V^{(m,n)}_{\mu_1\cdots\mu_n}\sigma^m\omega^{\mu_1}\cdots\omega^{\mu_n}\right) +\Sigma_{\rm s}=0. 
\label{eq:E20}
\\
{\delta \Gamma\over{\delta \omega^\mu}}
&=&-{\partial\over{\partial \omega^\mu}}\left(\sum_{m+n^>_=2}\int V^{(m,n)}_{\mu_1\cdots\mu_n}\sigma^m\omega^{\mu_1}\cdots\omega^{\mu_n}\right) -\Sigma_\mu=0. 
\label{eq:E21}
\end{eqnarray}


\noindent
{\bf 3 Effective meson masses, effective couplings and EOS}

Because of the conditions (6) and (7), the quark self-energies are strongly related to the meson potential. 
Therefore, at finite temperature and density, 
the effective meson-quark couplings are strongly related to the effective meson masses. 
In the uniform and rotationally invariant matter, we obtain 
\begin{eqnarray}
{{m_\sigma^*}^2\over{{m_\sigma}^2}} \equiv {\partial^2 \epsilon \over{\partial\sigma^2}}
={g_{\rm s\sigma}^*\Pi g_{\rm s\sigma}^*\over{m_\sigma^2}}+{g_{\rm s\sigma}^*\over{g_\sigma}}~~~~~{\rm and}~~~~~~
{{m_\omega^*}^2\over{{m_\omega}^2}}\equiv -{\partial^2\epsilon \over{\partial\omega_0^2}}
=-{g_{\rm s\omega}^*\Pi g_{\rm s\omega}^*\over{m_\omega^2}}+{g_{\rm v\omega}^*\over{g_\omega}},
\label{eq:E154}
\end{eqnarray}
where $g_{\rm s\sigma}^*\equiv -{\partial \Sigma_{\rm s}\over{\partial \sigma}}$, $g_{\rm s\omega}^*\equiv -{\partial \Sigma_{\rm s}\over{\partial \omega_0}}$, $g_{\rm v\omega}^*\equiv -{\partial \Sigma_0\over{\partial \omega_0}}$ and $\Pi$ is the polarization function. 
The $m_\sigma$, $m_\omega$, $g_\sigma$ and the $g_\omega$ are the $\sigma$-meson mass, the $\omega$-meson mass, the $\sigma$-quark coupling and the $\omega$-quark coupling at zero temperature and zero density, respectively, and $\epsilon$ is the energy density of the system. 
If the effects of the mixing interaction, the term including $g_{\rm s\omega}^*$, can be neglected, the square of the effective $\omega$-meson mass is proportional to the effective $\omega$-quark coupling. 
Therefore, the effective $\omega$-quark coupling decreases as the effective $\omega$-meson mass decreases. 

In Fig. 1, we show the baryon density ($\rho_{\rm B}$) dependence of the binding energy of nuclear matter at zero temperature. 
In the calculation, we assume that $g_{\rm Ns \sigma}^*=3g^*_{\rm s\sigma}$ and $g_{\rm Nv \omega}^*=3g^*_{\rm v\omega}$, 
where $g_{\rm Ns \sigma}^*$ and $g_{\rm Nv \omega}^*$ are the effective $\sigma$-nucleon and $\omega$-nucleon couplings, respectively. 
In the nonlinear model (NLM) $g_{\rm Nv\omega}^*/g_{\rm N\omega} ={m_\omega^*}^2/m_\omega^2\sim 0.94$ at the normal density $\rho_{\rm B0}$, whereas $g_{\rm Nv\omega}^*/g_{\rm N\omega} ={m_\omega^*}^2/m_\omega^2=1$ in the linear model (LM). 
(See Fig. 2.) 
Although the effective $\omega$-meson mass decreases in the NLM, 
the EOS in the NLM becomes much softer than that in the LM. 

~

\noindent
{\bf 4 Summary}

In summary, starting from the many quark interaction, we have constructed the nonlinear $\sigma$-$\omega$ mdoel. 
The mesons are introduced as auxiliary fields. 
Effective quark-meson couplings are strongly related to effective meson masses, since they are derived simultaneously from the original many quark interactions. 
In this model, even if the effective $\omega$-meson mass decreases due to the partial chiral restoration, the effective $\omega$-quark (or nucleon) coupling decreases and the EOS of nuclear matter can become soft.

\begin{flushleft}

{\bf References}

[1] F. Weber, Gy. Wolf, T. Maruyama and S. Chiba, preprint nucl-th/0202071; 
C.H. Hyun, M.H. Kim and S.W. Hong, Nucl.Phys. {\bf A718} (2003) 709. 

[2] H. Kouno,  Y. Horinouchi and K. Tuchitani, Prog. Theor. Phys. {\bf 112} (2004) 831; K. Tuchitani et al., preprint, nucl-th/0407004. 

[3] 
See, e.g. T. Kashiwa and T. Sakaguchi, Phys. Rev. {\bf D68} (2003) 065002. 

[4] H. Reinhardt, Phys. Lett. {\bf B208} (1988) 15. 

[5] H. Kouno et al., in preparation. 

\end{flushleft}

\begin{center}

\begin{center}
\begin{tabular}{cc}
\includegraphics[height=80mm,width=64mm] {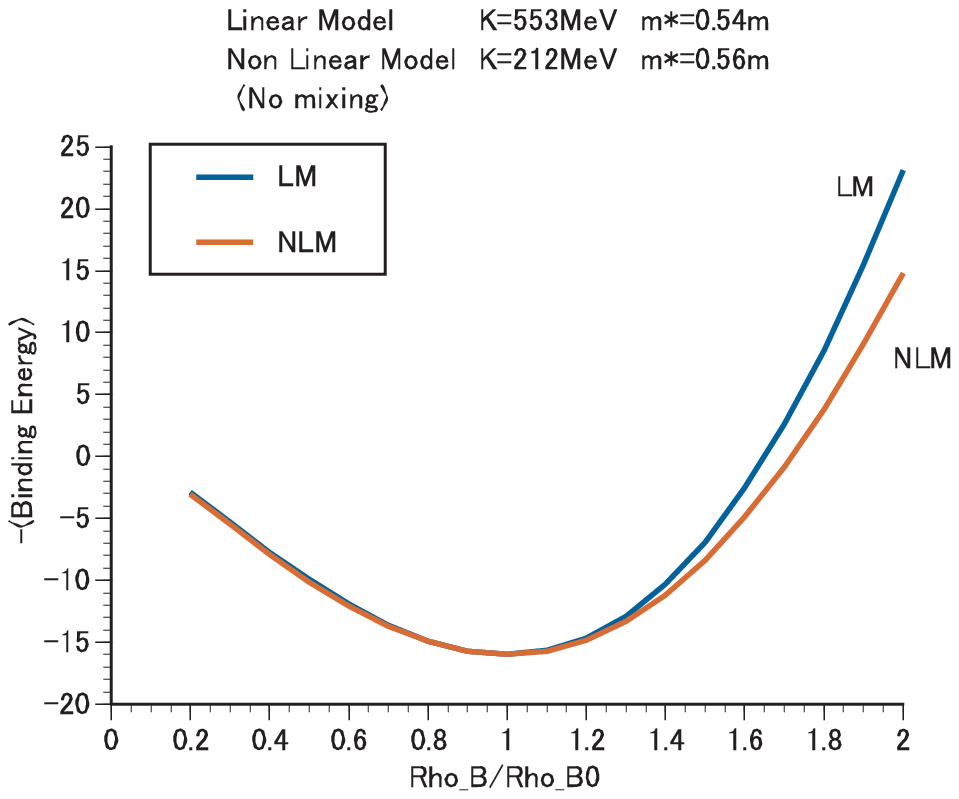} &
\includegraphics[height=80mm,width=64mm] {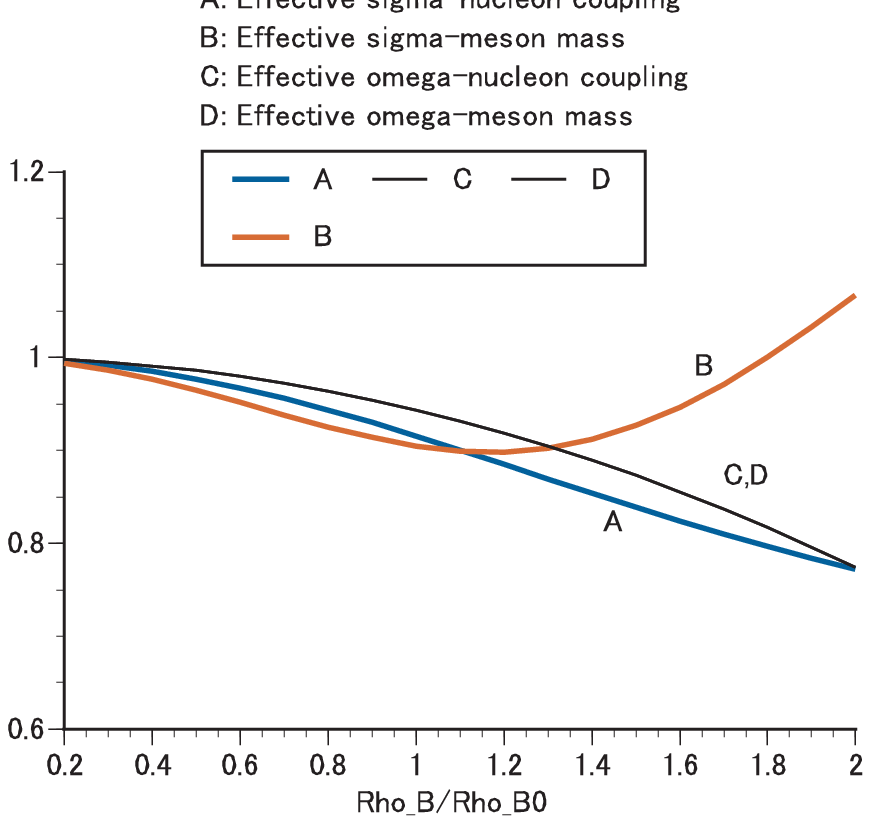}\\
{\bf Fig.1} Binding energy (in MeV) 
&
{\bf Fig.2} Squares of effective meson masses \\
$K$: Incompressibility & and effective meson-nucleon \\
$m^*$: Effective nucleon mass & couplings (ratios)  in NLM \\
at the normal density &     

\end{tabular}
\end{center}

\end{center}

\end{document}